# Measurements of Composite Fermion Conductivity Dependence on Carrier Density

C.-T. Liang[1], M.Y. Simmons[2,3], D.A. Ritchie[2], M. Pepper[2]

[1]*Department of Physics, National Taiwan University, Taipei 106, Taiwan*
[2]*Cavendish Laboratory, Madingley Road, Cambridge CB3 0HE, United Kingdom*
[3]*School of Physics, University of New South Wales, Sydney 2052, Australia*

We present the first experimental study of the carrier density dependence of the composite fermion conductivity $\sigma_{xx}^{CF}$ at Landau level filling factors $\nu = \frac{1}{2}$ and $\nu = \frac{3}{2}$ in high-quality front-gated GaAs/Al$_{0.33}$Ga$_{0.67}$As heterostructures. Extracting $\alpha$ from the power law $\ln(\sigma_{xx}^{CF}) \propto \ln(n_e)^{\alpha}$ shows that $\alpha \approx 1$. The measured $\alpha \approx 1$ is placed between the predicted value $\frac{3}{4}$ in the strong random magnetic field regime, and $\frac{3}{2}$ in the weak random magnetic field regime. Comparisons between our results and theory are discussed.

The fractional quantum Hall effect (FQHE) [1] observed in high-quality two-dimensional (2D) electron systems in the low-temperature, high magnetic field regime, arises from strong electron-electron interactions. These interactions cause the 2D electrons to condense into a fractional quantum Hall liquid [2]. In the elegant composite fermion (CF) picture [3], the FQHE can be understood as a manifestation of the integer quantum Hall effect of weakly interacting composite fermions. It has been shown that at a Landau level filling factor $\nu = 1/2$, a 2D electron system can be mathematically transformed into a composite fermion system interacting with a Chern-Simons gauge field [4,5]. At $\nu = 1/2$, the average of this Chern-Simons gauge field cancels the external magnetic field $B_{ext}$ so that the effective magnetic field acting on the CFs is zero. Away from $\nu = 1/2$, the composite fermions experience a net effective magnetic field $B_{eff} = (1 - 2\nu)B_{ext}$. To date, a wide variety of experimental results have supported the now well-established composite fermion picture [6–14].

One striking feature of composite fermions is that the resistivity $\rho_{xx}$ in the vicinity of $\nu = 1/2$ exhibits magneto-oscillations which look very similar to the familiar Shubnikov-de Haas oscillations around $B = 0$. However, there is one important difference–the oscillations in $\rho_{xx}$ near $\nu = 1/2$ are much more strongly damped than those at $B = 0$. Generally speaking, this is in line with the fact that $\rho_{xx}^{CF}$ at $B_{eff} = 0$ is orders of magnitude larger than $\rho_{xx}(B = 0)$ [15]. More specifically, recent theoretical studies have shown that the enhanced damping is due to scattering from fluctuations in the Chern-Simons gauge field as a result of the inhomogeneous electron charge distribution and resulting non-uniform screening of the impurity potential [4].

It is generally accepted that it is possible to determine the dominant scattering mechanism in a two-dimensional electron systems at $B = 0$ from the carrier density dependence of the mobility (conductivity) [16]. For example when $\mu \propto n_e^{1.5}$ then $\sigma \propto n_e^{2.5}$, since $\sigma = n_e e \mu$ and remote ionised impurity scattering dominates. However no such study has been conducted for composite fermions at $\nu = 1/2$, where the effective magnetic field for the CFs is also zero. A study of the carrier density dependence of the conductivity at $\nu = 1/2$ would determine what limits the CF mobility. To date, existing CF transport measurements have mostly been undertaken on un-gated, high-quality GaAs/AlGaAs heterostructures [6,9–11]. This paper presents the first study of the carrier density dependence of the composite fermion conductivity. In particular we show that the exponent $\alpha$ measured from the power law $\ln(\sigma_{xx}^{CF}) \propto \ln(n_e)^{\alpha}$ is approximately 1 at both $\nu = 1/2$ and $\nu = 3/2$. The measured $\alpha \approx 1$ is placed between the predicted value $\frac{3}{4}$ in the strong random magnetic field regime, and $\frac{3}{2}$ in the weak random magnetic field regime. Existing theories [4,15] underestimate $\sigma_{xx}^{CF}$ although the strong random magnetic field regime [15] provides a better fit to our results. We suggest that further theoretical studies are required in order to provide full understanding of our experimental results.

The high-quality HEMTs used in this work were made from ultra-low-disorder GaAs/Al$_{0.33}$Ga$_{0.67}$As heterojunctions. Sample A, made from wafer T139, has a carrier density of $9.12 \times 10^{14}$ m$^{-2}$ with a mobility $\mu$ of 300 m$^2$Vs$^{-1}$ at $V_g = 0$ without illumination. Sample B, made from wafer T205, has a 2DEG carrier density of $1.4 \times 10^{15}$ m$^{-2}$ and a mobility of 200 m$^2$Vs$^{-1}$ at $V_g = 0$ after brief illumination with a red light-emitting diode. Measurements were performed in a top-loading $^3$He cryostat at 0.3 K using standard four-terminal ac phase sensitive techniques.

Figure 1 (a) shows the four-terminal longitudinal resistivity $\rho_{xx}(B)$ for sample A at $V_g = 0$ V, showing high quality fractional states. Plotting the carrier density versus gate voltage in the inset shows there is a good linear fit over the measurement range $-0.3$ V $\leq V_g \leq 0$ V with $n_e = (2.14 \times 10^{15} V_g + 9.12 \times 10^{14})$ m$^{-2}$. The electron system is well described by a simple parallel plate capacitor model giving an estimated distance $D = 0.32$ $\mu$m between the front-gate and the underlying 2DEG. The calculated $D$ is in close agreement with the intended as-grown depth of 0.3 $\mu$m. Although the device could be operated in the accumulation mode, we concentrate on the case for $V_g \leq 0$ so that we could compare our results with existing theory as shown later. As shown in Fig. 1 (b), at the largest applied negative gate voltage $V_g = -0.3$ V we observe the fractional quantum Hall states $\nu = 1/3$ and $\nu = 2/3$, demonstrating the composite fermion picture is valid over the whole measurement range. At liquid helium temperatures, the distribution and density of the ionised impurities remain fixed, such that decreasing $V_g$ causes $n_e$ to decrease while the ionised impurity concentration $n_i$ remains constant.

We now present the main experimental finding of our paper. Figure 2 shows the composite fermion conductivity as a function of the carrier density for both samples at $\nu = 1/2$. At $\nu = 1/2$ each composite fermion is composed of an electron bound to two magnetic flux quanta [3], thus the density of the composite fermion system is equal to that of the electron system. A composite fermion can also be formed at $\nu = 3/2$. In this case, the spin-up level is filled whereas the spin-down level is half-filled. Therefore the composite fermion carrier density at $\nu = 3/2$ is one-third of the electron density [10]. Note that at $\nu = 1/2$ the composite fermion conductivity $\sigma_{xx}^{CF}$ is given by $1/\rho_{xx}(\nu = 1/2)$ whereas at $\nu = 3/2$, the composite fermion conductivity is given by $\frac{1}{9}$ of the inverse of the measured $\rho_{xx}$ [10]. Calculation of the exponent $\alpha$ from $\ln(\sigma_{xx}^{CF}) \propto \ln(n_e)^\alpha$ for Sample A and B are $1.02 \pm 0.06$ and $1.03 \pm 0.01$, respectively. The exponents for both samples at $\nu = 1/2$ are therefore very close to 1. Previously Coleridge, Zawadzki and Sachajda [17] reported a linear increase with increasing magnetic field of Shubnikov-de Haas peak values in the integer quantum Hall regime. However, their results are at various filling factors for a fixed carrier density. This is in sharp contrast to our experimental results on the composite fermion conductivity at different carrier densities.

Let us turn our attention to the case at $\nu = 3/2$. Figure 3 shows the composite fermion conductivity $\ln(\sigma_{xx}^{CF})$ as a function of the composite fermion carrier density $\ln(n_e/3)$ at $\nu = 3/2$ for both samples. Calculation of the slopes $\alpha$ for both fits gives exponents for Sample A and B as $0.96 \pm 0.04$ and $1.04 \pm 0.01$, respectively. From figures 2 and 3 it is evident that the slope of $\ln(\sigma_{xx}^{CF})$ versus $\ln(n_e)$ is $\approx 1$ *both* at $\nu = 1/2$ and $\nu = 3/2$. This experimental finding clearly establishes a link between $\nu = 1/2$ and $\nu = 3/2$ – suggesting that the carrier density dependence of the composite fermion conductivity is *the same* in both regimes. We also find $\sigma_{xx}^{CF}(\nu = 3/2) \approx (0.39 \pm 0.02)\sigma_{xx}^{CF}(\nu = 1/2)$ and $\sigma_{xx}^{CF}(\nu = 3/2) \approx (0.389 \pm 0.002)\sigma_{xx}^{CF}(\nu = 1/2)$ for Sample A and B, respectively. This is consistent with the fact that the effective disorder within the system at $\nu = 3/2$ is higher than that at $\nu = 1/2$ since the electrons occupied in the spin-up level are unable to screen the disorder [10].

Various analogies between the behaviour of the conductivity at $B=0$ and $\nu = \frac{1}{2}$ have been reported in the literature, such as on geometric resonances [8] and magnetic focussing [12]. In figure 4 we investigate the relation between the conductivity versus carrier density by plotting $\ln(\sigma_{xx})$ as a function of $\ln(n_e)$ at zero magnetic field and compare this with the data obtained at $\nu = 1/2$. At $B=0$ the measured exponents $\alpha$ are $1.82 \pm 0.02$ and $1.74 \pm 0.01$ which corresponds to $\mu \propto n_e^{0.82 \pm 0.02}$ and $\mu \propto n_e^{0.74 \pm 0.01}$ for Sample A and Sample B respectively. These results highlight that at B=0 background impurity scattering limits the mobility in our samples in agreement with previous studies of other high quality 2D electron systems [18,19]. However the value of $\alpha \approx 1.8$ obtained at B=0 is in sharp contrast to the value of $\alpha \approx 1$ obtained at $\nu = 1/2$ indicating that whatever limits CF mobility must be a different effect.

Seminal theoretical results in this field by Halperin, Lee and Read [4] suggest that in the CF regime additional scattering arises from the presence of random magnetic fields due to the inhomogeneous distribution of ionised dopants. In this work the composite fermion resistivity at $\nu = 1/2$ is given by $\rho_{xx} = \frac{n_i}{n_e} \frac{1}{k_F d_s} \frac{4\pi\hbar}{e^2}$, where $k_F = \sqrt{4\pi n_e}$ and $d_s$ is the spacer thickness. At $\nu = 1/2$, the composite fermion conductivity is given by the inverse of $\rho_{xx}$, such that

$$\sigma_{xx}^{CF}(\nu = \frac{1}{2}) = \frac{\sqrt{\pi} n_e^{\frac{3}{2}} d_s}{n_i} \frac{e^2}{h}. \quad (1)$$

In more recent work Mirlin, Polyakov and Wölfle (MPW) [15] have investigated the effects of disorder in a 2D system created by the non-uniform distribution of remote ionised impurities. They postulate that the inhomogeneous distribution of ionised dopants, of sheet density $n_i$ leads to fluctuations of the effective magnetic field $B(r) = (1 - 2\nu(r))B_{ext}$ (where $\nu(r)$ is the local filling factor) and thereby the Chern-Simons gauge field. It is the effect of a long-ranged random magnetic field (RMF) on CF transport that they model. There is a single parameter $\beta = \sqrt{\frac{n_i}{2n_e}}$ which determines the strength of the random magnetic field. For non-interacting CFs in a RMF Mirlin, Polyakov and Wölfle consider that there are two limits: the weak RMF ($\beta \ll 1$) and the strong RMF regimes ($\beta \gg 1$). In the weak RMF regime, they show that the composite fermion conductivity at zero $\bar{B}$ is given by

$$\sigma_{xx}^{CF}(\nu = \frac{1}{2}) = \frac{k_F d_s}{4\beta^2} = \frac{\sqrt{\pi} n_e^{\frac{3}{2}} d_s}{n_i} \frac{e^2}{h}, \quad (2)$$

where $k_F = \sqrt{4\pi n_e}$ and $d_s$ is the correlation radius (the spacer thickness) respectively. We can see that Eq. 1 is identical to Eq. 2 and predicts an exponent $\alpha = \frac{3}{2}$ in the relation $\ln(\sigma_{xx}^{CF}) \propto \ln(n_e)^\alpha$. However Mirlin and co-workers note that experimentally it is difficult to reach the weak RMF regime since the CF mean free path $l$ is rather short ($\approx 1$ $\mu$m) and therefore the theoretical assumption $l \gg d$ does not hold. In the strong RMF-regime MPW show that

$$\sigma_{xx}^{CF}(\nu = \frac{1}{2}) \approx \frac{k_F d_s}{\sqrt{\beta}} = \frac{2^{\frac{5}{4}} \pi d_s n_e^{\frac{3}{4}}}{n_i^{\frac{1}{4}}} \frac{e^2}{h}, \quad (3)$$

where $\alpha = \frac{3}{4}$ in the relation $\ln(\sigma_{xx}^{CF}) \propto \ln(n_e)^\alpha$ is predicted. There is an extra logarithmic factor in a detailed derivation of the formula (see Eq. 16 in Ref. [20])

In our system, as with the models of HLR and MPW, we assume that at zero gate voltage $V_g = 0$, $n_i = n_e$ and that $n_e$ decreases with decreasing $V_g$ while $n_i$ remains constant. For the range of carrier densities studied we calculate $\beta$ to be between 0.71 and 2.6 for Sample A and between 0.71 and 1.15 for Sample B. Thus our samples are placed between the weak $1 \gg \beta$ RMF regime and strong $\beta \gg 1$ RMF regime. Figure 5(a) and (b) show $\sigma_{xx}^{CF}$ as a function of carrier density $n_e$ for samples A and B, together with the theoretical curves for both limits of the model. We can see that both limits underestimate $\sigma_{xx}^{CF}$ for our system although the strong RMF field regime provides a better fit. The measured $\alpha \approx 1$ from the relation $\ln(\sigma_{xx}^{CF}) \propto \ln(n_e)^\alpha$ is placed between $\frac{3}{2}$ (weak RMF regime $1 \gg \beta$) and $\frac{3}{4}$ (strong RMF regime $\beta \gg 1$). However since we have $0.71 \leq \beta \leq 2.6$ in our system, we know we are in the crossover region between the strong and weak RMF regimes [20] ($0.2 \leq \beta \leq 10$) and this may explain the intermediate value of the measured exponent $\alpha$.

Another possible reason as to why $\alpha$ lies in the intermediate regime is the following. Assuming that all the donors are ionised ($n_i = n_e$ at $V_g = 0$) then $\beta \geq 0.71$ and can no longer be described in the weak RMF regime where $1 \gg \beta$. Therefore we can only consider the strong RMF regime which corresponds to $\beta \geq 10$. In this case, the corresponding electron densities of sample A and sample B are $4.5 \times 10^8$ m$^{-2}$ and $6.9 \times 10^8$ m$^{-2}$, respectively. At such low electron densities, the disorder within the electron system prohibits the observation of the fractional quantum Hall effects and the composite fermion picture is *no longer* valid.

It is worth mentioning that the composite fermion picture is not the only understanding of the FQHE. Extending the pioneering work of Laughlin [2], Haldane [21] and Halperin [22] proposed a hierarchical scheme to explain the FQH states observed in experiments. In their approach, low-lying excitations (quasi electrons or quasi holes) carry fractional charge. Within this picture, the $\nu = \frac{1}{2}$ state could be regarded as charge-neutral quasi particles at the infinite hierarchy level subject to a finite magnetic field. It may be possible that within the hierarchy picture, one would be able to develop a theory consistent with our experimental results of $\alpha = 1$.

In conclusion, we have measured the carrier density dependence of the composite fermion conductivity at $\nu = 1/2$ and $\nu = 3/2$ and show that the measured exponents $\alpha$ from the relation $\ln(\sigma_{xx}^{CF}) \propto \ln(n_e)^\alpha$ are found to be *both* $\approx 1$ in both cases. This suggests a link between the behaviour at $\nu = 3/2$ and at $\nu = 1/2$, despite the differing amounts of effective disorder present. Our results are in line with the enhanced scattering in the composite Fermion regime recently predicted by Mirlin, Polyakov and Wölfle [15] due to the random magnetic fields arising from the non uniform distribution of ionised donors. The precise value of $\alpha$ obtained experimentally was found to be in between the theoretical values required for either the weak $\frac{3}{2}$ or strong $\frac{3}{4}$ random magnetic field regimes recently proposed by Mirlin, Polyakov and Wölfle [15]. Both existing theoretical models [4,15] underestimate the composite fermion conductivity in our system although the theory in the strong random magnetic field regime [15] provides a better fit to our data. We suggest that in order to provide full understanding of our experimental results on $\alpha = 1$, further theoretical studies are required.

This work was funded by the NSC, Taiwan (grant no: NSC 89-2112-M-002-052, NSC 89-2112-002-084 and NSC 89-2911-I-002-104) and the EPSRC (UK). We would like to thank A.D. Mirlin for drawing our attention to Ref [15,20] and many fruitful discussions. C.T.L. is grateful to financial support from the Research Council, National Taiwan University, and Y.-M. Cheng, T.-Y. Huang, and C.G. Smith for experimental help.

[1] D.C. Tsui, H.L. Stormer and A.C. Gossard, Phys. Rev. Lett. **48**, 1559 (1982).
[2] R.B. Laughlin, Phys. Rev. Lett. **50**, 1395 (1983).
[3] J.K. Jain, Phys. Rev. Lett. **63**, 199 (1989).
[4] B.I. Halperin, P.A. Lee and N. Read, Phys. Rev. B **47**, 7312 (1993).
[5] A. Lopez and E. Fradkin, Phys. Rev. B **44**, 5246 (1991).
[6] R.R. Du *et al*, Phys. Rev. Lett. **70** 2944 (1993).
[7] R.L. Willett *et al*, Phys. Rev. Lett. **71**, 3846 (1993).
[8] W. Kang *et al*, Phys. Rev. Lett. **71**, 3850 (1993).
[9] D.R. Leadley *et al*, Phys. Rev. Lett. **72**, 1906 (1994).
[10] L.P. Rokhinson, B. Su and V.J. Goldman, Phys. Rev. B **52**, R11588 (1995).
[11] W. Kang *et al*, Phys. Rev. Lett. **75**, 4106 (1995).
[12] J.H. Smet *et al*, Phys. Rev. Lett. **77**, 2272 (1996).
[13] C.-T. Liang *et al*, Phys. Rev. B **53**, R7596 (1996).
[14] I.V. Kukushkin, K. von Klitzing and K. Eberl, Phys. Rev. Lett **82**, 3665 (1999).
[15] A.D. Mirlin, D.G. Polyakov and P. Wölfle, Phys. Rev. Lett. **80**, 2429(1998).
[16] For a review, see J.J. Harris, J.A. Pals and R. Woltjer, Rep. Prog. Phys. **52**, 1217 (1989).
[17] P.T. Colerige, P. Zawadzki and A.S. Sachrajda, Phys. Rev. B **49**, 10798 (1994).
[18] L. Pfeiffer *et al*, Appl. Phys. Lett. **55**, 1888 (1989).
[19] V. Umansky, R. de-Picciotto and M. Heiblum, Appl. Phys. Lett. **71**, 683 (1997).
[20] F. Ever, A.D. Mirlin, D.G. Polyakov and P. Wölfle, Phys. Rev. B **60**, 8951 (1999).
[21] F.D.M. Haldane, Phys. Rev. Lett. **51**, 605 (1983).
[22] B.I. Halperin, Phys. Rev. Lett. **52** 1583 (1984).

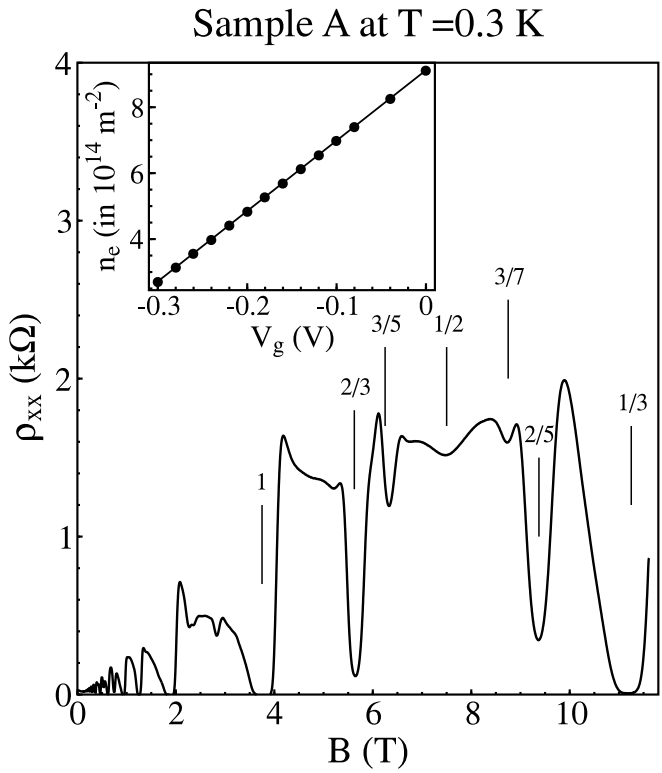
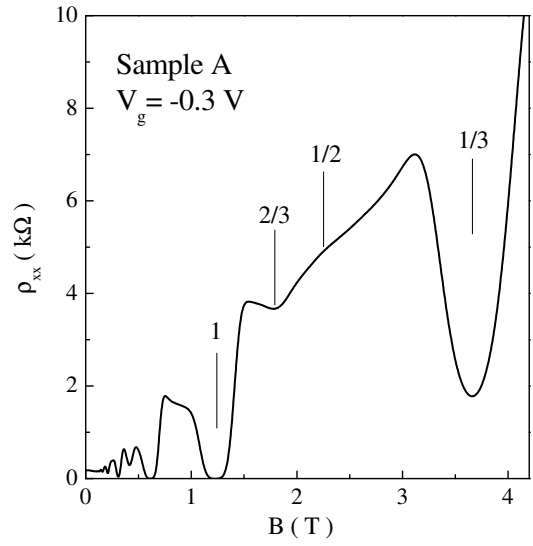

FIG. 1. (a) Magnetoresistivity measurements $\rho_{xx}(B)$ at $V_g = 0$ V for Sample A. The inset shows the carrier density as a function of applied front-gate voltage $V_g$. The linear fit is discussed in the text. (b) Magnetoresistivity measurements $\rho_{xx}(B)$ at $V_g = -0.3$ V for Sample A.

Figure 2 Liang et al.

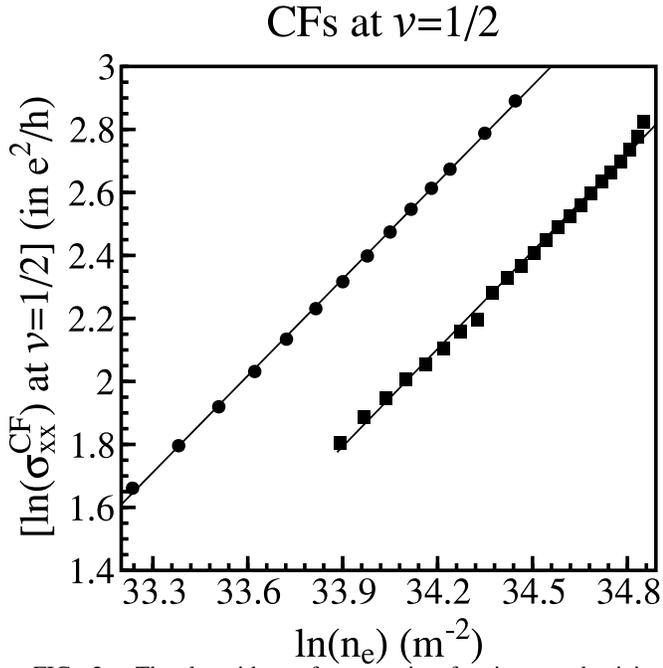

FIG. 2. The logarithm of composite fermion conductivity $\ln(\sigma_{xx}^{CF})$ as a function of the carrier density $\ln(n_e)$ at $\nu = 1/2$ for Sample A (marked by circles) and Sample B (marked by squares). The slopes of linear fits $\alpha$ for Sample A and B are $1.02 \pm 0.06$ and $1.03 \pm 0.01$, respectively.

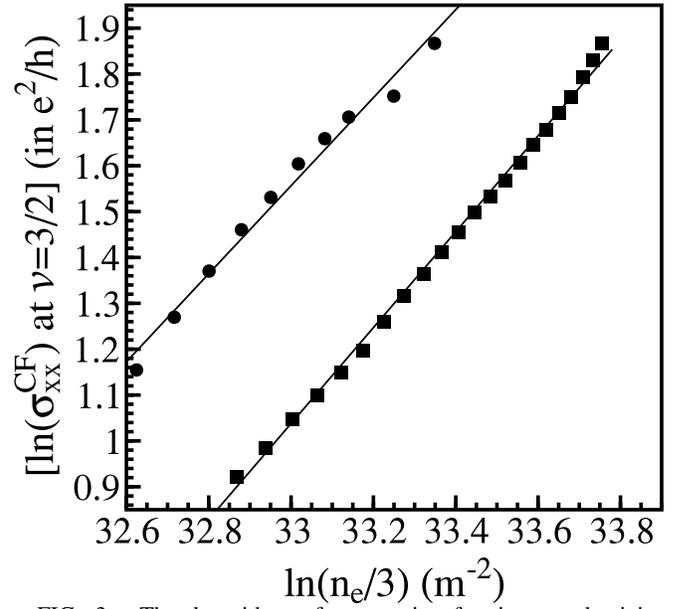

FIG. 3. The logarithm of composite fermion conductivity $\ln(\sigma_{xx}^{CF})$ as a function of carrier density $\ln(n_e)$ at $\nu = 3/2$ for Sample A (marked by circles) and Sample B (marked by squares). The slopes of linear fits $\alpha$ for Sample A and B are $0.96 \pm 0.04$ and $1.04 \pm 0.01$, respectively.

Figure 4 Liang et al.

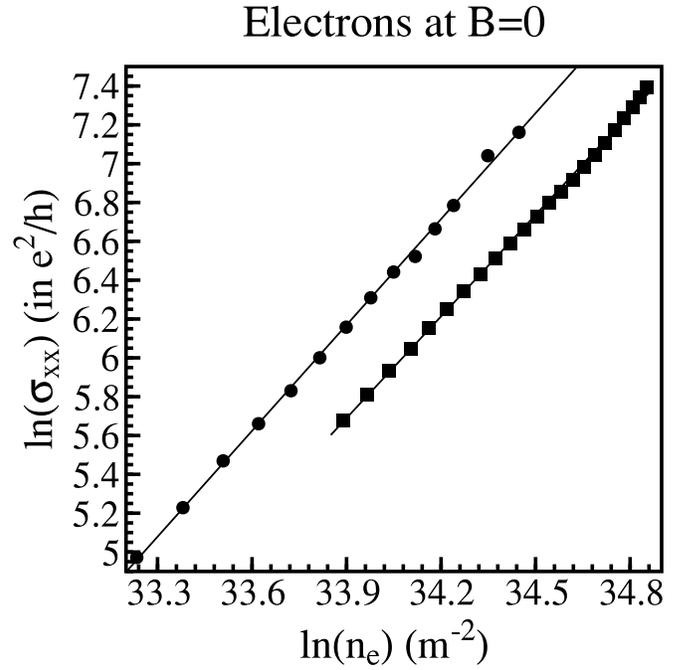

FIG. 4. The logarithm of electron conductivity $\ln(\sigma_{xx})$ as a function of carrier density $\ln(n_e)$ for Sample A (marked by circles) and Sample B (marked by squares) at $B = 0$. The slopes of the linear fits $\alpha$ for Sample A and Sample B are $1.82 \pm 0.02$ and $1.74 \pm 0.01$, respectively

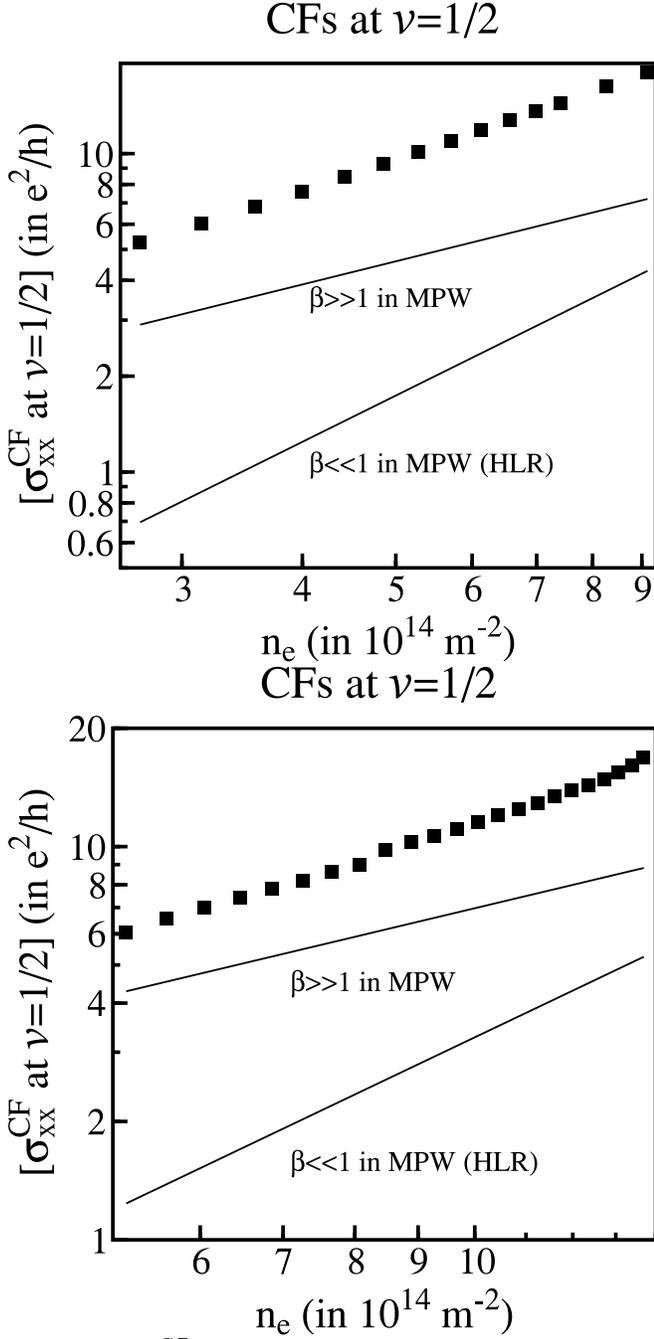

FIG. 5. (a) $\sigma_{xx}^{CF}$ as a function of carrier density $n_e$ at $\nu = 1/2$ for Sample A. (b) $\sigma_{xx}^{CF}$ as a function of carrier density $n_e$ at $\nu = 1/2$ for Sample B. The solid lines represent the limits of the random magnetic field model as discussed in the text.